\documentclass{PoS}

\newcommand{\be}{\begin{equation}}
\newcommand{\ee}{\end{equation}}
\newcommand{\bary}{\begin{eqnarray}}
\newcommand{\eary}{\end{eqnarray}}

\usepackage{lineno}

\title{Search of extended or delayed TeV emission from GRBs with HAWC}

\ShortTitle{Extended or delayed emission from GRBs with HAWC}

\author{\speaker{Simone Dichiara}$^{a}$, Maria Magdalena Gonz\'alez$^{a}$, Nissim Fraija$^{a}$, Ibrahim Torres$^{b}$, Ana Delia Becerril$^{c}$, Ruben Alfaro$^{c}$, Dirk Lennarz$^{d}$ for the HAWC collaboration$^{e}$\\
        \llap{$^a$} Instituto de Astronom\'{i}a, Universidad Nacional Aut\'{o}noma de M\'{e}xico, M\'{e}xico D.F., M\'{e}xico\\
        \llap{$^b$} Instituto Nacional de Astrof\'{i}sica, \'{O}ptica y Electr\'{o}nica, Puebla, M\'{e}xico\\
        \llap{$^c$} Instituto de F\'{i}sica, Universidad Nacional Aut\'{o}noma de M\'{e}xico, M\'{e}xico D.F., M\'{e}xico\\
        \llap{$^d$} School of Physics and Center for Relativistic Astrophysics, Georgia Institute of Technology, Atlanta, Georgia, USA\\
        \llap{$^e$}For a complete author list, see \href{http://www.hawc-observatory.org/collaboration/icrc2017.php}{http://www.hawc-observatory.org/collaboration/icrc2017.php}.\\
        
        E-mail: \email{sdichiara@astro.unam.mx}}

\abstract{Gamma-ray bursts (GRBs) are among the most luminous sources in the universe and the nature of their emission up to very high energy is one of the most important open issue connected with the study of these peculiar events. The High Altitude Water Cherenkov (HAWC) gamma-ray observatory, installed at an altitude of 4100 m a. s. l. in the state of Puebla (Mexico), has completed its second year of full operations. Thanks to its instantaneous field of view of $\sim$ 2 sr and its high duty cycle ($\geq$ 95\%), HAWC is an ideal instrument for the study of transient phenomena such as GRBs. We performed a search for TeV emission delayed with respect to, and of longer duration than the prompt emission observed by satellites. We present here the results obtained by observing at the position of a sample of GRBs detected by the Fermi and Swift satellites from December 2014 to February 2017. The upper limits resulting from this analysis are presented and theoretical implications are discussed. }

\FullConference{35th International Cosmic Ray Conference --- ICRC217\\
		10--20 July, 2017\\
		Bexco, Busan, Korea}

\begin{document}

\section{Introduction}

After 50 years from their first discovery from the Vela satellites in 1967 \cite{bib:klebesadel73}, the physical mechanism at the origin of gamma-ray bursts (GRBs) is far from being completely understood. Several models are proposed to explain the different phases of their emission. 
The standard scenario proposed to explain these phenomena consists of a fireball in which the energy is emitted by internal and external shocks \cite{bib:rees94, bib:narayan92}. The prompt emission is produced by dissipation of kinetic energy driven by internal collisions (see \cite{bib:piran04} for a review).  The energy spectrum is highly non-thermal and usually described by an empirical function (so-called Band function \cite{bib:band93}). 
The late emission, so-called afterglow, is visible up to days or weeks after the explosion and arises when the flow is slowed down by collisions with the surrounding circumburst medium (either a stellar wind of its progenitor or interstellar medium, ISM) \cite{bib:fraija2017}. 
Based on the duration of their prompt emission and their hardness ratio (the ratio between the total number of counts detected at high and low energies), GRBs are mainly associated with two groups; long and short GRBs \cite{bib:kouveliotou93}. Long GRBs (lGRBs) have a $T_{90}$ longer than 2 seconds and the short GRBs (sGRBs) have a $T_{90}$ less than 2 seconds\footnote{$T_{90}$ is defined as the time during which the cumulative number of detected counts increases from 5\% to 95\% above background, thus encompassing 90\% of the total GRB counts}. SGRBs have typically higher hardness ratio with respect to the long class.
%

The Large Area Telescope instrument on board the Fermi satellite (Fermi/LAT, \cite{bib:atwood2009}) revealed a high-energy component (> 100 MeV) from over 100 GRBs\footnote{https://fermi.gsfc.nasa.gov/ssc/observations/types/grbs/lat\_grbs/}. At those energies the emission lasts more than that at lower energies and the spectra evolve differently. It is worth noting that the highest energy photon is often detected at late time with respect to the trigger (high energy photons are still observed several tens of seconds after the end of the prompt phase; e.g. \cite{bib:ackermann2011}).  
Several theoretical scenarios have been proposed to explain the origin of this component \cite{bib:fraija2015,bib:meszaros94}. High-energy radiation can be produced by inverse Compton upscattering of prompt and/or afterglow photons. 
In particular it can be interpreted as a jet running into an external medium (external shock model). Meszaros and Rees \cite{bib:meszaros94} suggested that the inverse Compton scattering could be the dominant cooling mechanism of accelerated electrons in the external shock wave when it gets close to the deceleration radius. 
The search and study of this high energy counterpart is a key point in the understanding of these phenomena and it is crucial to disentangle the complex jigsaw concerning the different zones of the emission. On the other side, it allows to probe phenomena such as the extra-galactic background light (EBL; see \cite{bib:schirber2003}), the initial Lorentz factor and Lorentz invariance violation (see e.g. \cite{bib:amelino1998}).
%

In this work, we will present the results obtained by the High Altitude Water Cherenkov observatory (HAWC) taking into consideration a sample of bursts in its field of view. The aim of this analysis is to detect high energy extended and/or delayed emission with respect to the prompt phase.

\section{HAWC}

HAWC is an extensive air shower array located at Sierra Negra, in the state of Puebla, Mexico. This observatory uses a water Cherenkov technique to study TeV $\gamma$-ray radiation and cosmic rays. The experiment comprises 300 tanks of 7.3 m diameter and 4.5 m depth, each filled with purified water. The full array covers an area of $\sim 2\times 10^4\, {\rm m}^{2}$. Each tank is equipped with 4 photomultiplier tubes (PMTs): a 10-inch PMT located at its center, and other three additional 8-inch PMTs distributed around it. HAWC scans $\sim 2/3$ of the sky every sidereal day. 
When a $\gamma$ photon interact with the atmosphere, it generates an extensive air shower of relativistic particles that reaches the detector plane. The particles interact with the water in the tanks and produce optical light via Cherenkov radiation. Measuring the arrival time and the time over threshold of all the PMT pulses in each tank, we reconstruct the core of the shower and the direction of the primary particle. Studying the distribution of the charge on the detector array, we can select the gamma-ray photons using a set of specific gamma-hadron (GH) separation cuts (see \cite{bib:abeysekara2017} for more details). 
The events detected by HAWC are classified in ten different size bins (from 0 to 9) which are based on the fraction of available PMT activated by the event as described by \cite{bib:abeysekara2017}. The lowest size bin (bin 0) is excluded from this analysis since no final set of GH separation cuts is available for these events. 
%
%
%
Previous works already analyzed the HAWC data looking for possible emission from GRBs (e.g. \cite{bib:lennarz2015,bib:lennarz2017}) although none of them reports any significant detection. 

\section{Data Analysis}

For this analysis we consider all the GRBs observed by HAWC from December 2014 to  February 2017.  We analyze the position of 93 GRBs observed by the two instruments of Fermi, LAT (from 20 MeV to 300 GeV) and GBM (Gamma-Ray Burst Monitor; from 10 to 1000 keV), and by BAT (the Burst Alert Telescope; from 15 to 150 keV) on board Swift. 

During this period, 7 bursts observed by Fermi/LAT were in the field of view of HAWC, 6 of them were also detected by the Fermi/GBM and 3 were co-detected by Swift/BAT (GRB 150314A, GRB 160821A and 161202A). 
For all the common cases between Fermi and Swift we keep the position provided by Swift, since it is the most accurate one. 
Out of the three aforementioned GRBs, Swift detected other 27 GRBs in the field of view of HAWC during the same period of time. Seventy-seven GRB positions were provided only by Fermi/GBM. 
Eighteen GRBs are excluded from the analysis due to issues with the data as explained in detail in \cite{bib:lennarz2017}. Out of the ones presented in \cite{bib:lennarz2017} we exclude GRB 160821A (detected by Fermi/LAT) since it occurred during a calibration run and two Fermi/GBM GRBs (160824598 and 160827616) because HAWC was out of power during these days. We also ignore GRB 150716A (detected by Swift/BAT) and GRB 160131174 (detected by Fermi/GBM) since HAWC did not collect enough data to study the signal at late time; therefore they are not suitable for the purpose of this work. 

Three different methods are currently used to inspect the HAWC data looking for possible emission coming from GRBs: 1) focusing the search around the GRB trigger time (from $-T_{90}$ to $2 \times T_{90}$) and looking for the emission using a sliding window mode; 2) checking the signal over three time windows of different duration ($T_{90}$, $3 \times T_{90}$, $10 \times T_{90}$ for lGRBs and $T_{90}$, 6 seconds and 20 seconds for sGRBs); and 3) measuring the significance in ten consecutive time windows of duration of $T_{90}$ and 2 seconds for lGRBs and sGRBs, respectively.
We run simulations to test the efficiency of the different approaches, concluding that the third method would be the most sensitive for the detection of possible extended or delayed (late time) emission.

\section{Results}

\begin{figure}[htbp]
\centering
\includegraphics[width=.7\textwidth]{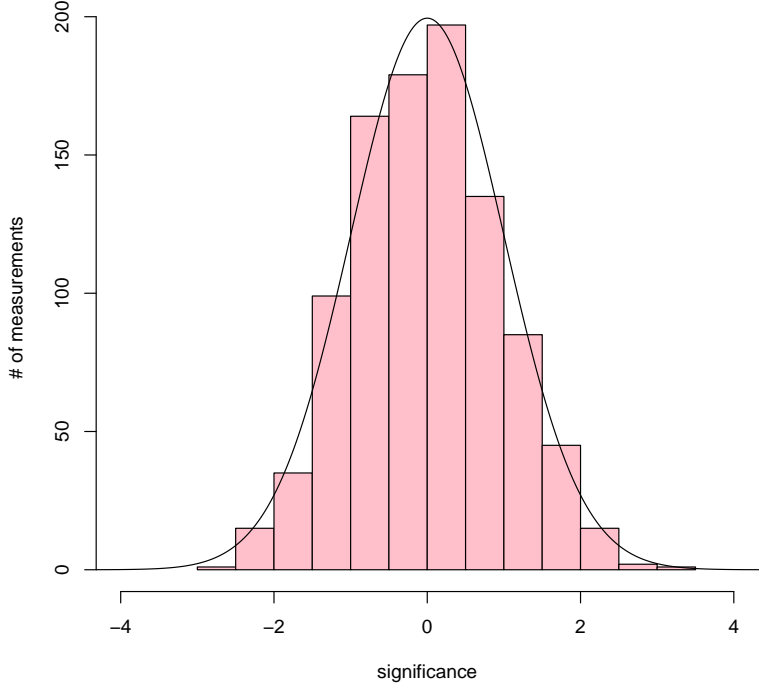}
\caption{The distribution of the significances derived over 10 different time windows for each GRB of our sample. The results fit well with a Gaussian function centered at 0 and standard deviation of 1 (solid line).}
\label{significance_distribution}
\end{figure}

The highest value for the significance is $3.07 \sigma$ obtained for the GBM GRB bn151023104 in the time interval between $40.96$ and $51.2$ seconds from the trigger time ($0.34 \sigma$ considering the trials correction). The distribution of the significances extracted in each time windows for all the GRBs is displayed in Fig.~\ref{significance_distribution}. It does not allow us to claim the presence of any signal in the data, neither at early nor late times.
Upper limits for the number of events are derived using the method described by \cite{bib:feldman1998} following the same procedure outlined in \cite{bib:lennarz2017}. We use Monte Carlo simulations (see \cite{bib:abeysekara2017}) to derive the HAWC effective area and convert these values in upper limits for the fluxes and the fluences in the energy range $80-800$ GeV. For this computation, we assume a power law spectrum with an index -2 and we consider the attenuation effects due to the EBL according to the fiducial model proposed by \cite{bib:gilmore2012}. Considering that the bulk of sGRB redshift measurements are between 0.1 and 1.3, with average value around 0.5 (see \cite{bib:berger2014}), we assume a conservative value of z=1.0 
whenever the actual measure is not available. The most constraining upper limits are derived for GRB 170206A. 

\subsection{GRB 170206A}
\begin{figure}[htbp]
\centering
\includegraphics[width=.7\textwidth]{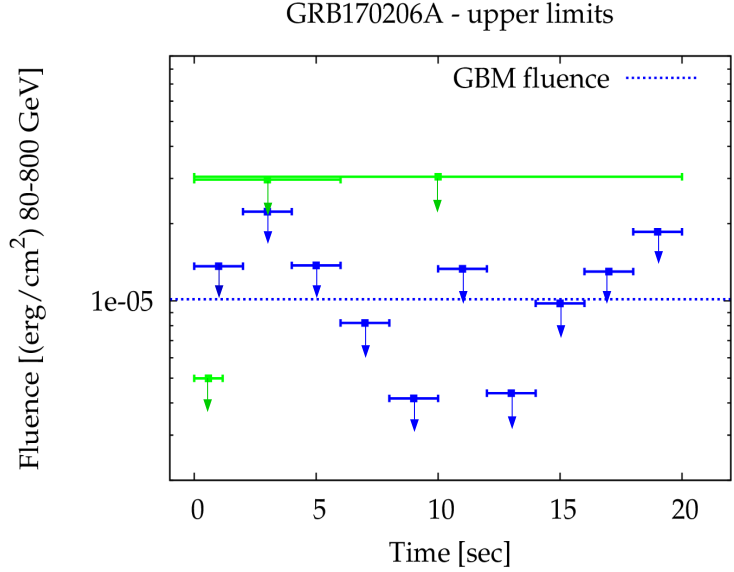}
\caption{Arrows show the 170206A fluence upper limits extracted in the energy range 80-800 GeV over different time windows and considering a redshift of 1. Blue and green arrows show the limits extracted over ten consecutive windows of 2 seconds and over three windows of different duration ($T_{90}$, 6 seconds and 20 seconds), respectively. Green limits are shown to compare our results with the ones presented in \cite{bib:lennarz2017}. The dashed line indicates the fluence measured by Fermi-GBM in the energy range 10-1000 keV}
\label{170206A_upper_limits}
\end{figure}

GRB 170206A is the third brightest short burst detected by Fermi-GBM. This is a very interesting case since HAWC is more sensitive to short and bright signals (see \cite{bib:taboada2014}). Fluence upper limits derived for this burst over different time windows are shown in Fig.~\ref{170206A_upper_limits}. This figure shows the upper limits derived for the fluence between $80-800$ GeV assuming z=1.0. We used time windows of 2 seconds to extract the significances and compute the upper limits.

\subsubsection{Model}

\begin{figure}[htbp]
\centering
\includegraphics[width=.7\textwidth]{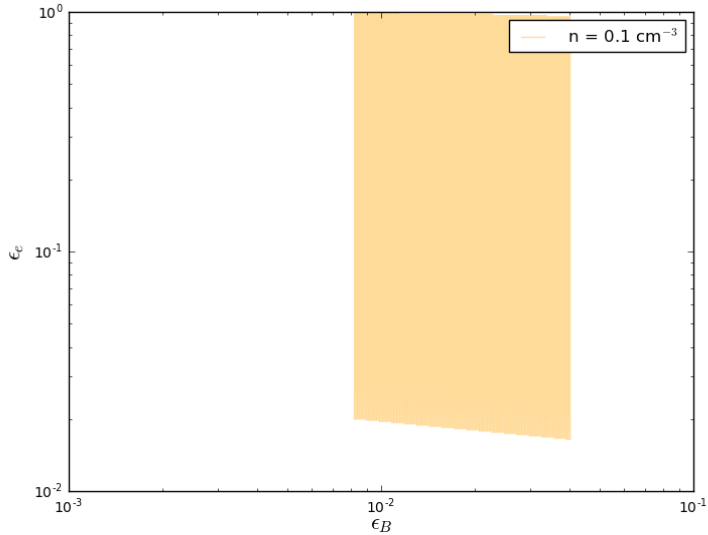}
\caption{The allowed range for the microphysical parameters $\epsilon_B$ and $\epsilon_e$ derived using the GRB 170602A flux upper limits. We assume a density of 0.1 particles per $\mathrm{cm}^{3}$}
\label{microphysics}
\end{figure}

In the external shock model \cite{bib:fraija2016b}, afterglow emission is generated when expanding relativistic ejecta encounter the circumburst medium and, consequently, forward and reverse shocks are produced.  Relativistic electrons are accelerated at the shocks to a power law distribution by the first order Fermi mechanism  $\gamma_e\geq\gamma_m: N(\gamma_e)\,d\gamma_e \propto \gamma_e^{-p}\,d\gamma_e$,  and cooled down by synchrotron and inverse Compton (synchrotron self-Compton; hereafter SSC) scattering emission.  Here, $\gamma_m$ is the minimum Lorentz factor. Comparing  the time scales of  cooling time,  the deceleration time  and the acceleration time of electrons, the synchrotron spectral breaks as a function of time $t$ are:
{\small
\bary\label{synfor_b}
E^{\rm syn}_{\rm \gamma,m}&\propto& \epsilon_{e}^2\,\epsilon^{1/2}_{B}\,E^{1/2}\,t^{-3/2}\,,\cr
E^{\rm syn}_{\rm \gamma,c}&\propto&  \epsilon^{-3/2}_{B}\,n^{-1/2}\, E^{-1/2}\,t^{-1/2}\,,\cr
E^{\rm syn}_{\rm \gamma,max}&\propto&  n^{-1/8}\,E^{1/8}\,t^{-3/8}\,,
\eary
}
where $n$ is the homogenous density,  $E$ is the  the isotropic equivalent kinetic energy,  $\epsilon_e$ and $\epsilon_B$ are the microphysical parameters defined as fraction of the energy density that goes to accelerate electrons and amplify the magnetic field, respectively. The synchrotron spectral breaks  $E^{\rm syn}_{\rm \gamma,m}$, $E^{\rm syn}_{\rm \gamma,c}$, $E^{\rm syn}_{\rm \gamma,max}$ are the characteristic, the cooling and the maximum break energies, respectively.  The maximum synchrotron flux can be written as
{\small
\bary\label{Fsyn_for}
F^{\rm syn}_{\rm \gamma,max}\propto \epsilon^{1/2}_{B}\,n^{1/2}\,D^{-2}\,E\,,
\eary
}
where $D$ is the luminosity distance.   Accelerated electrons can upscatter photons from low to high energies as {\small $E^{ssc}_{\gamma, m/c}\simeq \gamma_e^{2}  E^{syn}_{\gamma, m/c}$}. The break SSC energies can be written as
{\small
\bary\label{sscf}
E^{\rm ssc}_{\rm \gamma,m}&\propto& \,\epsilon_e^{4}\,\epsilon_{B}^{1/2}\,n^{-1/4}\,E^{3/4}\,t^{-9/4}\cr
E^{\rm ssc}_{\rm \gamma,c}&\propto&\,\epsilon_{B}^{-7/2}\,n^{-9/4}\,E^{-5/4}\,t^{-1/4}\,,
\eary
}
and the maximum SSC flux is
\bary
F^{\rm ssc}_{\rm \gamma,max}\propto \epsilon_{B}^{1/2}\,n^{1/4}\,D^{-2}\,E^{5/4}\,t^{1/4}\,.
\eary
Klein-Nishina (KN) correction to the spectrum is important in the high-energy part.  The emissivity is reduced compared with the classical inverse Compton regime which is given by {$\small E^{\rm KN}_{\rm \gamma,c,f}\propto\epsilon_{B,f}^{-1}\,n^{-3/4}\,E^{-1/4}\,t^{-1/4}$}.  Considering the case $E^{\rm ssc}_{\rm \gamma,c} \leq E_\gamma \leq E^{\rm ssc}_{\rm \gamma,m}$, the flux as a function of time in the corresponding energy regime is
\be
F^{\rm ssc}_{\rm \nu} \propto \epsilon_{B,f}^{-5/4}\,n^{1/8}\,D^{-2}\,E^{5/8}\,t^{1/8}\, E_\gamma^{-1/2} 
\ee
Flux upper limits extracted at different times (see Section 4) can be used to constrain the normalization of eq. 4.5, thus the allowed range for the microphysical parameters $\epsilon_B$ and $\epsilon_e$ assuming different values for n (and $\Gamma$). Fig.~\ref{microphysics} shows the allowed values for $\epsilon_B$ and $\epsilon_e$ derived assuming n = 0.1 $\mathrm{cm}^{-1}$ and considering the upper limits measured for GRB 170206A. In this case $\epsilon_B$ can range between $\sim 8\times10^{-3}$ and $\sim 4\times10^{-2}$ while $\epsilon_e$ ranges between $\sim 2\times10^{-2}$ and $\sim 1$. These values are consistent with the ones previously reported in the literature for other GRBs \cite{bib:santana2014,bib:berger2014}. Moreover, our results indicate that the ISM density must be lower than 0.8 particles per $\mathrm{cm}^{3}$ to be the model consistent with the measured limits.  

\section{Discussion and conclusions}

We used the HAWC data to search for VHE emission at the position of 93 GRBs detected between December 2014 and February 2017. We focused our search on possible delayed or extended emission by extracting the signal significance over different time windows for each case. Although our analysis does not allow us to claim any evidence of GRB emission from the data, constraining upper limits for the flux coming from GRB 170206A are derived. Taking into account eq. (4.5), we can use these limits to constrain the microphysical parameters and to probe the prediction proposed for the VHE emission in the framework of the external shock model. We find that the density of the ISM is n$<0.8\, {\mathrm cm^{-3}}$ and the relativistic Lorentz factor $\Gamma=\biggl(\frac{3}{32\,\pi\,m_p}\biggr)^{1/8}\,(1+z)^{3/8}\,n^{-1/8}\,E^{1/8}\,t^{-3/8}\,$ > 800, where $m_p$ is the proton mass. 

In this work we derived upper limits from simultaneous VHE observations of GRBs and used them to constrain the microphysical parameters of the SSC mechanism in the fast cooling regime. 
The detection of new constraining upper limits, together with a multiwavelength analysis would provide us with an important tool to test physics of the GRB afterglow and to probe the nature of the VHE emission. 

\section*{Acknowledgments}
\footnotesize{
We acknowledge the support from: the US National Science Foundation (NSF); the US Department of Energy Office of High-Energy Physics; the Laboratory Directed Research and Development (LDRD) program of Los Alamos National Laboratory; Consejo Nacional de Ciencia y Tecnolog\'{\i}a (CONACyT), M{\'e}xico (grants 271051, 232656, 260378, 179588, 239762, 254964, 271737, 258865, 243290, 132197), Laboratorio Nacional HAWC de rayos gamma; L'OREAL Fellowship for Women in Science 2014; Red HAWC, M{\'e}xico; DGAPA-UNAM (grants RG100414,IN111315, IN111716-3, IA102715, 109916, IA102917); VIEP-BUAP; PIFI 2012, 2013, PROFOCIE 2014, 2015;the University of Wisconsin Alumni Research Foundation; the Institute of Geophysics, Planetary Physics, and Signatures at Los Alamos National Laboratory; Polish Science Centre grant DEC-2014/13/B/ST9/945; Coordinaci{\'o}n de la Investigaci{\'o}n Cient\'{\i}fica de la Universidad Michoacana. Thanks to Luciano D\'{\i}az and Eduardo Murrieta for technical support
}

\bibliography{HAWC_GRB_proceeding_v0}
%

\end{document}